\documentclass[onecolumn,showpacs,preprintnumbers,amsmath,amssymb]{revtex4}
\usepackage{graphicx}
\usepackage{dcolumn}
\usepackage{bm}
\usepackage{color}
\usepackage{amsmath}
\usepackage{xcolor}
\usepackage{caption}
\usepackage{float}
\usepackage{soul}
\newcommand{\kalpha}{K$_\alpha$\,}
\newcommand{\kbeta}{K$_\beta$\,}

\begin{document}

\title{Laser-based X-ray and electron source for X-ray fluorescence studies}

\author{F. Valle Brozas\,$^{1}$}\email{fvalle@clpu.es}
\author{A. Crego\,$^2$} \email{acrego@usal.es}
\author{L. Roso\,$^1$}\email{roso@clpu.es}
\author{A. Peralta Conde\,$^1$}\email{aperalta@clpu.es}

\affiliation{$^1$ Centro de L\'aseres Pulsados, CLPU, Parque Cient\'ifico, 37185 Villamayor, Salamanca, Spain.\\
$^2$ Facultad de Ciencias, Universidad de Salamanca, 37008 Salamanca, Spain.}
\date{\today}

\begin{abstract}
In this work we present a modification to conventional X-rays fluorescence using electrons as excitation source, and compare it with the traditional X-ray excitation for the study of pigments. For this purpose we have constructed a laser-based source capable to produce X-rays as well as electrons. Because of the large penetration depth of X-rays, the collected fluorescence signal is a combination of several material layers of the artwork under study. However electrons are stopped in the first layers allowing therefore a more superficial analysis. We show that the combination of both excitation sources can provide extremely valuable information about the structure of the artwork.
\end{abstract}


\maketitle

\section{Introduction}

The incessant development of laser technology in the last decades providing systems with a growing energy per pulse, shorter pulses, and extremely robust with respect to the experimental parameters, has paved the way to new and unexpected applications in several scientific disciplines. In fact, one can say, without fear of being deceived, that laser technology development  has become a leading force for the progression of new laser-based tools that get advantage of all these new capabilities. One of the fields that has been clearly reinforce by this progress is laser-based particle acceleration. In simple terms the acceleration of particles is achieved by focusing an ultraintense laser pulse with a pulse duration of the order of the central wavelength $\rm \lambda$, i.e., pulses that contain a limited number of optical cycles, in a spot of the order of $\rm \lambda^2$.  In such conditions the laser energy is concentrated in a spatial cube of the order of $\rm \lambda^3$ being possible to achieve intensities up to 10$^{17}$\,Wcm$^{-2}$ with moderate laser powers (GW).  In these acceleration mechanisms, the key factor is the achieved intensity in the target point. For intensities of the order of 10$^{16}$\,Wcm$^{-2}$ the electric field of the laser overcomes the electric field that the electrons feel from the nucleus and the atom is ionized. The expeled electrons are further accelerated by the laser, and re-injected into the bulk material once the electric field changes its direction. In this process deep ultraviolet (VUV), extreme ultraviolet (XUV), and X-ray radiation are generated  (see for example \cite{Krausz06, Pukhov06, Lefebvre13} and references therein).  For intensities of 10$^{18}$\,Wcm$^{-2}$ not only electrons are accelerated but also protons, and if the laser intensity increases to 10$^{20}$\,Wcm$^{-2}$ there are mechanisms for the acceleration of neutrons (see \cite{Passoni13, Pirozhkov12} and references therein). 

In the following we will concentrate in the production of X-rays and electrons with moderate intensities 10$^{16}$-10$^{17}$\,Wcm$^{-2}$. In simple terms, this generation process can be understood as follow. When the laser interacts with the target, as the intensity is considerable high, in the early stages of the pulse the matter gets ionized. Then, the rest of the pulse interacts rather than with an electrically neutral target but with an expanding plasma from the target surface. Accordingly the description of the process must be done in terms of laser-plasma interaction. Since the intensity is still sufficient, the laser extracts electrons from this plasma, accelerates, and re-injects them into the bulk producing both Bremsstrahlung radiation by the sudden loss of energy of the re-injected electrons, and the characteristics X-ray emission of the material of the target. It is important to notice that the laser electric field does not only re-inject electrons into the target but also accelerates them in the direction of the laser reflection. According to this, this source is capable of providing X-ray pulses and bunches of electrons both with a temporal duration of the order of the laser pulse.

The kind of sources described previously have attracted a large attention lately because of their versatility, as well as their reduced size and price when compared to conventional particle accelerator.  One possible application of these sources not explored so far to the best of our knowledge, is the use of a laser-based X rays and electron source as a tool for non-destructive analysis of artworks.  X-ray fluorescence (XRF) is a well established technique that provides valuable information about the presence of chemical elements in a sample (see for example \cite{XRF1, XRF2, XRF3} and references therein).  When the sample is irradiated with X-rays (or electrons), inner electrons of the atoms of the samples are excited. Once the produced vacancies in the electronic configuration are occupied by other electrons, the atoms emits a characteristic X-ray emission that is a fingerprint of each element. Nowadays the importance of XRF is beyond doubt. For example when applied to paintings, this technique reveals the elemental composition of the pigments used by the artist helping the art restorers to prevent the degradation of the colors as well as to answer questions related with authenticity and provenance. However the interpretation of the XRF data is not always straightforward. X-ray emission from the instrument itself and/or the surroundings of the artwork under analysis, e.g., the holder, can induce misinterpretation. Another important factor that must be taken into account is the penetration depth of the X-rays. For example considering that oil paintings have as a basis a drying oil, e.g., poppy seed oil or safflower oil, with densities lower than 1\,gcm$^{-3}$ a material like PMMA with a density of 1.18\,gcm$^{-3}$ can be used to determine an upper limit for the X-ray penetration. According to \cite{nist_attenuation} 10\,keV X-rays are attenuated up to 90\,\% in 0.6\,cm of PMMA, while for 100\,keV this thickness is of around 12\,cm. Thus, one must be aware that the obtained X-ray fluorescence is produced not only on the superficial paint layer, but also on those behind the visible one. This difficulty can be overcome if electrons are used as excitation source. 

In this manuscript we describe our results using a laser-based of X-rays and electrons source as excitation for XRF analysis.  The use of these allows us to study not only several layers of pigment simultaneously using X-rays as in conventional XRF technique, but also due to the limited penetration depth of electrons we can study just the surface of the artwork.  In the following we will first briefly describe our sources providing a characterization for the generated X-rays and electrons. Then we will present the X-ray fluorescence of different pigments obtained by exciting them with X-rays and electrons. We will discuss the experimental data underlying the differences between both excitation sources. Finally, we complete our contribution with a brief summary and an outlook.

\section{Setup}

\begin{figure}
\includegraphics[width=0.5\columnwidth]{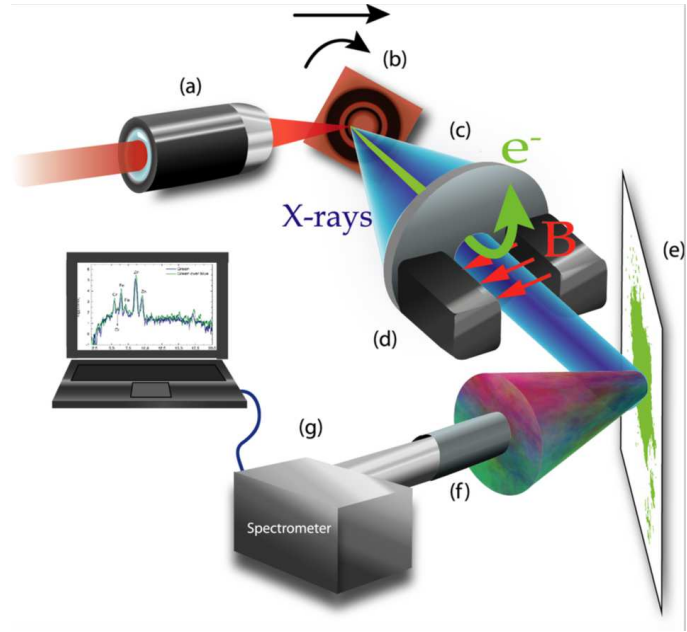}
\caption{\label{fig1} (Color online) Experimental setup: (a) microscope objective (b) cooper target (c) collimator (d) magnets (e) sample (f) collimator (g) spectrometer.}
\end{figure}

For the production of X-rays and electron bunches a Titanium-Sapphire femtosecond laser system with a pulse length of 120\,fs, operating at 1\,kHz repetition rate, a carrier wavelength of 800\,nm, and a circular spatial profile with a radius of 0.6\,cm FWHM (Full Width Half Maximum) was used. The laser system can provide up to 7.5\,mJ but for the experiments discussed in this work an energy per pulse of just 0.9\,mJ was enough for the generation of X-rays and electrons. This energy was focalized in the target by a microscope objective of numerical aperture NA = 0.42 (see \ref{fig1}). The achieved focal spot measured at low intensity was in the order of $1.5\,\mu$m in the horizontal and $1.2\,\mu$m in the vertical direction.  With this spot size the expected intensity is of $\rm \sim5\cdot$10$^{17}$\,Wcm$^{-2}$.  However, we must notice that at high energy conditions the focal spot will be considerable larger due to filamentation (see for example \cite{Chin10} and reference therein). For this situation we estimate a focal spot with a diameter of the order of 100\,$\mu$m and an intensity of $\rm \sim1\cdot$10$^{14}$\,Wcm$^{-2}$. It is noticeable that even at such low intensities X-ray and electron bunches are produced.  For ensuring a fresh target surface for each laser shot, the target was continuously spinning and moving perpendicular to the laser (see Fig.\,\ref{fig1}). Without these movements the laser focus quality will be damage because the material is ablated for every shot. It is interesting to notice that although for this work we have used cooper as a target, it is possible to used any material, e.g., plastics, mylar, or PMMA. For the deflection of the electrons from the X-ray emission direction we placed a collimator with an aperture of around 1\,cm$^2$ and a pair of magnets (see Fig.\,\ref{fig1}).  To verify the direction of the electron bunches once deflected we used gafchromic films (EBT2). The X-ray fluorescence of the different samples was collected by an Amptek Silicon drift detector (SDD-132) placed close to them.

\begin{figure}
\includegraphics[width=0.5\columnwidth]{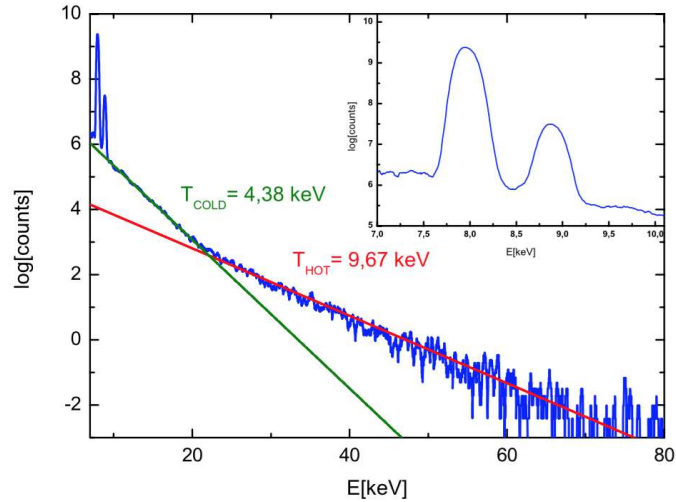}
\caption{\label{fig2} (Color online)  X-ray spectrum obtained from a cooper target. The Bremsstrahlung emission has its origin in an electron energy spectrum described by a bi-component Maxwell-Boltzmann distribution with temperatures T$\rm _{hot}$=9.67\,keV and T$\rm _{cold}$=4.38\,keV. The characteristics \kalpha and \kbeta lines of cooper with energies of 8.0\,keV and 8.9\,keV are also visible.}
\end{figure}

Figure\,\ref{fig2} shows a typical spectrum of the generated X-rays using cooper as a target. We can clearly see two distinctive features. On one hand the Bremsstrahlung emission produced by those electrons extracted from the target by the laser, accelerated by the electric field, and finally re-injected. This emission has two different origins: the electrons directly accelerated by the laser and the electrons accelerated by collisions mechanisms within the plasma during the plasma expansion. As a result of it, the electrons energy spectrum is described by a bi-component Maxwell-Boltzmann distribution. We can estimate the temperature of these components by adjusting the produced Bremsstrahlung X-rays spectrum resulting T$\rm _{hot}$=9.67\,keV and T$\rm_{cold}$=4.38\,keV respectively \cite{Gibbon07, Brozas}. On top of the Bremsstrahlung emission we can see the characteristics \kalpha and \kbeta lines of the target material, Cu in this work, at 8.0\,keV and 8.9\,keV respectively \cite{nist}. When the accelerated electrons are re-injected into the plasma, they have acquired enough energy for ionizing the atoms of the target material. More concretely, via electron impact ionization an electron from an inner shell is promoted into the vacuum level producing the characteristic X-ray emission when the electron vacant is filled by other electrons. It is important to notice, that this dependence on the target material allows us to "tune" both the characteristic X-ray emission as well as the Bremsstrahlung. As a general rule, denser materials produce more energetic Bremsstrahlung emission than lighter ones.

As it was stated in the introduction, we can assume for X-rays of energy 10-100\,keV a penetration depth in oil paintings larger than 0.6-12\,cm respectively. For electrons we can proceed in a similar way. If we consider electrons of 20\,keV the stopping power for PMMA is 12.8\,MeVcm$^2$g$^{-1}$ \cite{nist2}. Thus, the electrons penetrate up to 18\,$\mu$m in the material

\section{Experimental results and discussion}

\begin{figure}
\includegraphics[width=0.5\columnwidth]{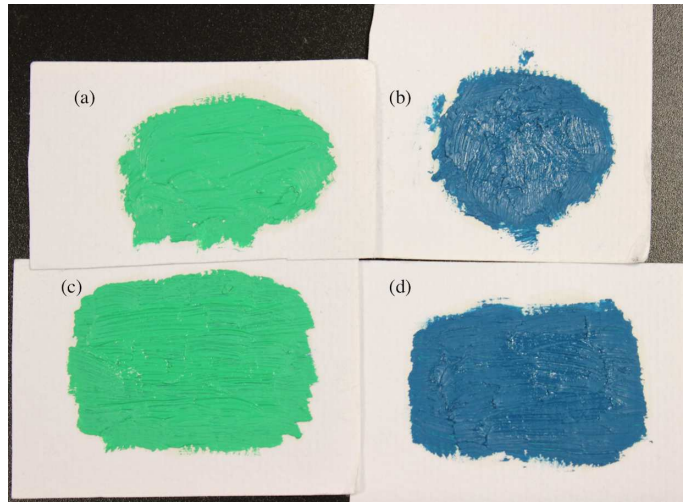}
\caption{\label{fig3} (Color online) Samples used for the measurements. We used as pigments emerald green and cerulean blue over watercolor paper: (a) solely green, (b) solely blue (c), green over blue, and (d) blue over green.}
\end{figure}

For obtaining the experimental results discussed in this section we proceed as follows. We used as pigments emerald green and cerulean blue (Oil pigments Marie's, usual composition Cu(CH$_3$COO)$_2\cdot$3Cu(AsO$_2$)$_2$ and CoO$\cdot$nSnO$_2$  \cite{composition}) over watercolor paper in four different samples: green and blue solely, green over blue, and blue over green (see Fig.\,\ref{fig3}). In a first set of measurements to have reference spectra we exposed directly the samples to the electrons and X-ray radiation. For this, we removed the deflecting magnets shown in Fig.\,\ref{fig1}. Also we obtained a spectrum without any sample to calibrate the X-ray emission of the materials of the setup.Then, we measured the composed samples with X-rays and electrons, comparing the results with the spectra of reference. 

\begin{figure}
\includegraphics[width=0.5\columnwidth]{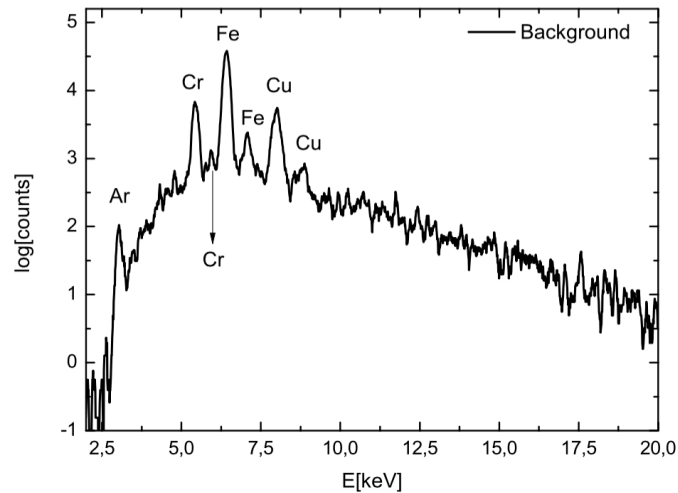}
\caption{\label{fig4} (Color online) Background spectrum.}
\end{figure}

Figure\,\ref{fig4} shows a spectrum obtained without sample. This measurement allowed us to know the background produced mainly by the X-ray fluorescence of the materials the source is made of, i.e., post holders, optical mounts, etc. We identified the \kalpha and \kbeta emissions of the following elements: Cr 5.4\,keV and 5.9\,keV, Fe 6.4\,keV and 7.1\,keV, Ni 7.5\,keV and 8.3\,keV, and Cu 8.0\,keV and 8.9\,keV \cite{nist}. All these materials were present in the experimental setup. At 2.7\,keV we can see the emission lines of the Ar of air (natural abundance 0.934\,\%). Argon is the only specie of air with emission lines energetic enough to be observed in our detector (lower limit of 1\,keV). Since this peak is present with the same magnitude for all the samples, we have normalized all spectra to it.

\begin{figure}
\includegraphics[width=0.5\columnwidth]{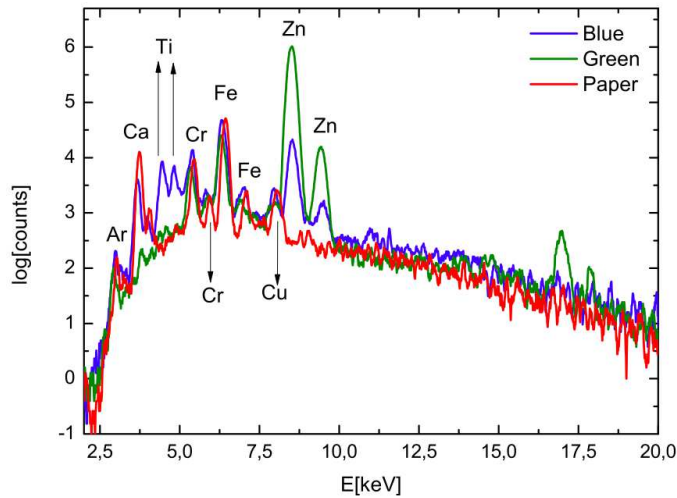}
\caption{\label{fig5} (Color online) Reference X-ray fluorescence spectrum for the blue and green pigments, and the watercolor paper.}
\end{figure}

Figure\,\ref{fig5} shows the obtained XRF results for the blue and green pigments (see Fig.\,\ref{fig3}\,(a) and (b)), as well as for the watercolor paper. The structure of the obtained spectra is very rich containing peaks from the background already identified in Fig.\,\ref{fig4}, the pigments, and the watercolor paper. Analyzing Fig.\,\ref{fig5} from low to high energies, the first unidentified peaks lie at 3.7\,keV and 4.0\,keV. We assigned these emissions to the \kalpha and \kbeta lines of Ca. Our hypothesis is that Ca is present in the form of calcium carbonate (CaCO$_3$). This compound is normally used in the pigment industry as a whitening agent. Interestingly for the green pigment this element was not identified although we used the same watercolor paper for both (blue and green) samples. Different hypothesis can be proposed to explain this fact. Likely the green pigment has a strong absorption at these energies preventing the detection of the Ca lines. For the blue pigment either it does not absorb these energies, or it absorbs it but the pigment has been whitened with a Ca compound.

Continuing the analysis of Fig\,\ref{fig5}, we clearly identified for the blue pigment at 4.5\,keV and 4.9\,keV the \kalpha and \kbeta lines of Ti. Titanium is normally used in the pigment industry in the form of titanium dioxide (TiO$_2$). This compound is usually known as titanium white. At 8.5\,keV and 9.5\,keV are located the \kalpha and \kbeta lines of Zn. This element has been detected for both pigments, notice that it was not present in the watercolor paper, although for the green concentration is significantly higher than for the blue. Zinc is normally presents as zinc oxide (ZnO) and it is usually known as zinc white. Finally, at 17\,keV and 17.9\,keV no reasonable element assignation could be made. According to this, we attributed these signals to pile-up events of the Zn lines.  Pile-up happens when the detector is not capable to discriminate between two events that are close in time. In this case, rather than two events the detector emits a signal corresponding to a single event but with double the energy. Unfortunately this is a common problem for particle and radiation spectrometers that becomes specially relevant for pulse radiation due to high instantaneous fluxes achieved. Nowadays due to the fast development of laser-based ultrashort particle and radiation sources there is a growing interest in this instrumental artifact. 

According to the chemical composition of the pigments, it was expected to find the emission of Co and Cu for the blue and green colors respectively. In the first case, the \kalpha emission of Co lies at 6.9\,keV which it is close to the much more intense \kbeta line of Fe located at 7.1\,keV from the background. For Cu the situation is similar. There is a strong background emission from the setup and the target itself, hiding probably the Cu emission of the pigment. It is also possible that the amount of these element necessary to generate the color is tiny, being therefore below our detection limit. Alternatively it is plausible that the color was generated adding organic pigments like for example indigo C$_{16}$H$_{10}$N$_2$O$_2$ whose x-ray emission lie below 1\,keV and are normally not detected by XRF techniques.

\begin{figure}
\includegraphics[width=0.5\columnwidth]{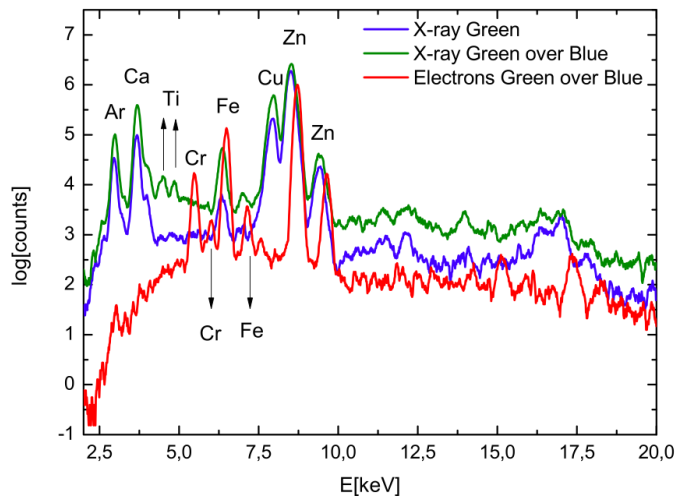}
\caption{\label{fig6} (Color online) X-ray fluorescence spectrum for the green pigment when excited by X-rays, and for the composed sample green over blue when excited by X-rays and electrons.}
\end{figure}

Figure\,\ref{fig6} shows the results obtained from a green layer of paint over a blue one (see Fig.\,\ref{fig3}\,(c)) when it was excited by X-rays and electrons separately. For comparison purposes a XRF spectrum of solely the green pigment when excited by X-rays is also shown. It must be mentioned that although X-rays measurements are normalized to the Ar signal as it was discussed above, for the electron measurements this is not possible because the experimental setup was different. Let us focus at the Ti lines at 4.5\,keV and 4.9\,keV. As it was discussed previously this element is present in the cerulean blue but not in the emerald green pigment (see Fig.\,\ref{fig5}). However for the composed sample green over blue, one can easily identified the signature of this element. The penetration depth of X-rays at these energies ($\sim$600\,$\mu$m for 10\,keV) is sufficient for producing the fluorescence not only from the upper paint layer but also from the lower one. Thus the obtained fluorescence signal is a combination of different layers of paints complicating therefore the identification of the pigments. This difficulty can be overcome if electrons are used as excitation source as it is shown in Fig.\,\ref{fig6}. Since the penetration depth of electrons is much lower ($\sim 20\,\mu$m for 20\,keV electrons) only the superficial layer is excited, and consequently the measured spectrum does not have any contribution of the blue color underneath.

\begin{figure}
\includegraphics[width=0.5\columnwidth]{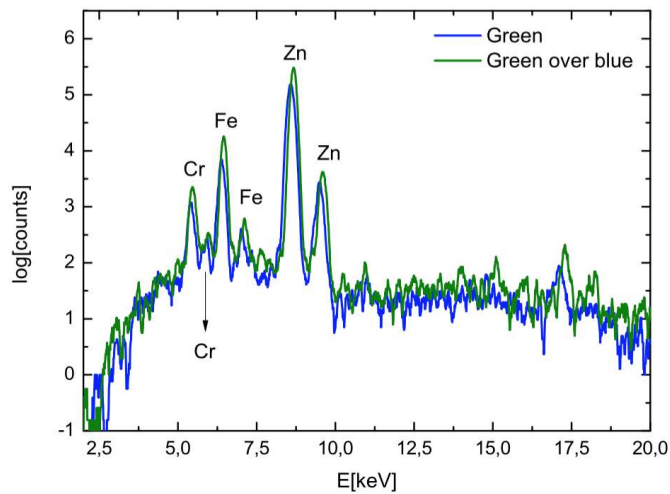}
\caption{\label{fig7} (Color online) X-ray fluorescence spectrum of the green pigment and the composed sample green over blue when excited by X-rays.}
\end{figure}

To obtain further confirmation of the results discussed above, we collected the X-ray fluorescence of green and green over blue when exciting by electrons. As it is shown in Fig.\,\ref{fig7} both curves are essentially identical not showing any evidence of the blue pigment under the green one. Specifically there is no signature of Ti as it can clearly see in Fig.\,\ref{fig6}. This clearly indicated that when complex samples are excited by electrons only the superficial layer produces an appreciable fluorescence signal. 

To finalize the analysis of Fig\,\ref{fig6} it must be said that in contrast to the measurements shown in Fig.\,\ref{fig5} the Ca lines at 3.7\,keV and 4.0\,keV are visible. We attribute this to the slightly larger collimator used for these measurements. The same argument applies to the large Cu peak at 8.9\,keV (compare Fig.\,\ref{fig5} and Fig.\,\ref{fig6}). This modified setup allowed us to increase the signal but, as a drawback, part of the watercolor paper without pigments was also excited.

\begin{figure}
\includegraphics[width=0.5\columnwidth]{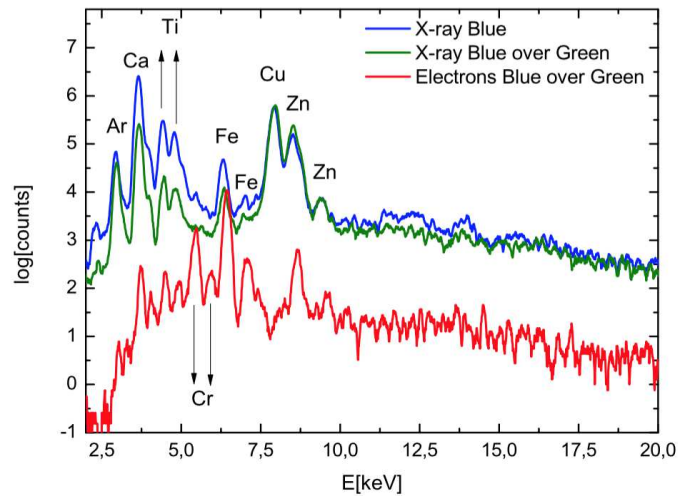}
\caption{\label{fig8} (Color online) X-ray fluorescence spectrum for the blue pigment when excited by X-rays, and for the composed sample blue over green when excited by X-rays and electrons.}
\end{figure}

Figure\,\ref{fig8} shows the obtained results when a blue over green composite sample (see Fig.\,\ref{fig3}\,(d)) was irradiated by X-rays and electrons. In this case the interpretation is more complex than for the previous sample (green over blue) because of the richer fluorescence spectrum of the blue pigment (see Fig.\,\ref{fig5}). Let us focus in the ratio between the Zn and Ti \kalpha lines located at 8.6 and 4.5\,keV respectively. As it can be seen in Fig.\,\ref{fig8} when the blue sample was irradiated by X-rays the fluorescence ratio between the Zn and Ti lines was 0.7. However when the composite sample blue over green was irradiated this ratio increased to 2.7, i.e., an increase of $\sim387\,\%$, because of the high concentration of Zn of the green pigment (see Fig.\,\ref{fig5}). When electrons were used as irradiation source because of their limited penetration depth this ratio was reduced to 1.2  which is only $\sim70\,\%$ larger than the original ratio. We attribute this $\sim70\,\%$ difference to the different excitation sources we have used, i.e., X-rays and electrons.  

\section{Conclusions}

In this work we have shown with a proof of principle experiment the potential of laser-based particles acceleration techniques in X-ray fluorescence analysis of complex samples. In particular we have shown the potential of the combine use of electrons and X-ray, both emitted simultaneously from our laser-based source, in carrying out XRF analysis of paint layers. While X-rays give us information about several paint layers as well as the canvas, electrons because of their limited penetration depth excite just the most superficial layer. According to this, using our source it is possible not only to simplify the analysis of the XRF spectra, but also to obtain extremely valuable information about the superficial layers of the artwork.   

\section{Acknowledgements}

The authors thank J.M. \'Alvarez for fruitful and stimulating discussions during the development of this manuscript. This work has been possible by the support from Ministerio de Educaci\'on Cultura y Deporte (F. Valle Brozas studentship FPU AP2012-3451) and Ministerio de Econom\'ia y Competitividad of Spain  (FURIAM project FIS2013-47741-R).

\end{document}